# The Computational Complexity of Sandpiles

Cristopher Moore[1] and Martin Nilsson[2]

[1] Santa Fe Institute, 1399 Hyde Park Road, Santa Fe, New Mexico 87501
moore@santafe.edu
[2] Chalmers Tekniska Högskola and University of Gothenburg, Göteborg, Sweden
martin@fy.chalmers.se

**Abstract.** Given an initial distribution of sand in an Abelian sandpile, what final state does it relax to after all possible avalanches have taken place? In $d \geq 3$, we show that this problem is **P**-complete, so that explicit simulation of the system is almost certainly necessary. We also show that the problem of determining whether a sandpile state is recurrent is **P**-complete in $d \geq 3$, and briefly discuss the problem of constructing the identity.

In $d = 1$, we give two algorithms for predicting the sandpile on a lattice of size $n$, both faster than explicit simulation: a serial one that runs in time $\mathcal{O}(n \log n)$, and a parallel one that runs in time $\mathcal{O}(\log^3 n)$, i.e. the class $\mathbf{NC}^3$. The latter is based on a more general problem we call Additive Ranked Generability. This leaves the two-dimensional case as an interesting open problem.

## 1 Introduction

One way to define the "complexity" of a physical system is the amount of time, memory, or other computational resources needed to predict it. How these resource requirements vary with the system size, or the number of time-steps we wish to predict, becomes increasingly important as we study larger and larger systems for longer and longer times.

For example, consider a cellular automaton (CA), a dynamical system with a discrete set of local states and a local update rule where each site's evolution depends on its state and those of its neighbors. We can predict the state of a given site $t$ time-steps in the future by explicitly simulating the CA and filling in a light-cone of depth $t$ ending in that site. This takes time $\mathcal{O}(t^{d+1})$ on a serial computer (proportional to the volume of space-time above the desired site) or $\mathcal{O}(t)$ on a massively parallel one.

However, for classes of CAs that obey certain algebraic identities, we can do much better [20, 21]. These "quasi-linear" systems can be predicted on a parallel computer in time $\mathcal{O}(\log t)$ or $\mathcal{O}(\log^2 t)$, much faster than by explicit simulation, placing them in the complexity classes $\mathbf{NC}^1$ or $\mathbf{NC}^2$. In general, $\mathbf{NC} = \cup_k \mathbf{NC}^k$ is the class of problems that can be solved in *polylogarithmic time* ($\mathcal{O}(\log^k n)$ for some $k$, where $n$ is the number of bits of the input) on a parallel computer with a polynomial number of processors. In this case, the input consists of $\mathcal{O}(t^d)$ sites of the initial state of the CA.

**NC** is a subset of **P**, the class of problems solvable in polynomial time on a serial computer [12, 26]. It is believed but not known that there are *inherently sequential* problems, in **P** but not in **NC**, where a substantial amount of work has to be done step-by-step order, and having a large number of processors gives at most a polynomial speedup.

**P**-*complete* problems, to which any other problem in **P** can be reduced, are the most likely to be inherently sequential. If any **P**-complete problem has a fast parallel algorithm, then every problem in **P** does, and **P** = **NC**. This would be highly surprising, and would violate our intuition that inherently sequential problems exist. (This is analogous to the better-known fact that if an **NP**-complete problem such as TRAVELING SALESMAN can be solved in polynomial time, then **NP** = **P**.)

Several problems of interest in physics have been shown to be **P**-complete by Machta, Moore and others, including Ising models [18, 23], diffusion-limited aggregation [16, 18], lattice gases [22], cellular automata with local voting rules [23], and simple deterministic growth models [13]. Greenlaw et al. [12] have pointed out that predicting a CA's evolution is **P**-complete in general, since CA rules exist (e.g. [15]) which can simulate universal Turing machines. On the other hand, **NC** algorithms exist for Eden growth [17] and the Lorentz lattice gas [24].

This distinction between **P** and **NC** is not just useful from a numerical point of view; it can be viewed as a fundamental fact about a system's dynamics. If a short-cut exists, then the system can be considered solvable to some extent. As an analogy, an integrable dynamical system with a closed-form solution can be predicted simply by plugging $t$ into an equation, while a non-integrable system may have to be integrated numerically. If explicit simulation is necessary, then the system is contingent on its history in an unavoidable way, and the relationship between initial and final states is deeply entangled.

In this paper, we study the SANDPILE PREDICTION problem, namely predicting the final state of an Abelian sandpile from an initial distribution of sand. We show that SANDPILE PREDICTION is **P**-complete in $d \geq 3$, as is the problem of determining whether a sandpile state is recurrent, and we briefly discuss the difficulty of constructing the identity state.

In $d = 1$, we give two fast algorithms for sandpile prediction, a serial one that runs in $\mathcal{O}(n \log n)$ time and a parallel one that runs in $\mathcal{O}(\log^3 n)$ time, showing that the one-dimensional problem is in $\mathbf{NC}^3$. Along the way, we discuss a more general problem we call ADDITIVE RANKED GENERABILITY, which is of some interest in its own right.

The $d = 2$ case remains an interesting open problem, and we discuss some possible approaches to a fast algorithm for it.

We note that related work on the computational power of sandpiles on general graphs and trees can be found in [3, 11], and related results on the one-dimensional sandpile can be found in [10].

## 2 Boolean circuits

The canonical **P**-complete problem is Circuit Value. Given a Boolean circuit, i.e. a directed graph whose nodes are AND, OR and NOT gates, and given the truth values of its inputs, is the output true or false? This is **P**-complete since any deterministic Turing machine computation of length $t$ can be converted to a Boolean circuit of depth $\mathcal{O}(t)$, so polynomial-time computations are equivalent to polynomial depth (and size) circuits.

Since the truth values at each level of the circuit affect those on the next level in nearly arbitrary ways, it is hard to imagine how one could calculate the output without going sequentially through the circuit level by level. An **NC** algorithm would have to somehow evaluate many levels at once, or provide a method of skipping over most of them. In fact, $\mathbf{P} = \mathbf{NC}$ if and only if circuits of polynomial depth always have much shallower equivalent circuits.

The Monotone Circuit Value problem, where AND and OR gates are allowed but NOT gates are not, is still **P**-complete for the following reason: using De Morgan's laws $\overline{a \wedge b} = \overline{a} \vee \overline{b}$ and $\overline{a \vee b} = \overline{a} \wedge \overline{b}$, we can shift negations back through the gates until they only affect the inputs themselves. Thus any Circuit Value problem can be converted to a Monotone Circuit Value problem with some of the inputs negated.

This kind of conversion from one problem to another is called a *reduction*. If $A$ can be reduced to $B$, then $B$ is at least as hard as $A$, since if a fast algorithm for $B$ exists, we can convert it to a fast algorithm for $A$ by first converting instances of $A$ to instances of $B$. The usual way to show that $B$ is **P**-complete is to first show that it is in **P**, and then to reduce some other **P**-complete problem to it.[3]

In [23], we show that majority-voting cellular automata and zero-temperature Ising models are **P**-complete by reducing Monotone Circuit Value to them. We do this by showing that the system's dynamics supports 'wires' that carry truth values, AND and OR gates that combine them, and ways to bend and split wires to connect them from one gate to another.

In three or more dimensions, we can then implement arbitrary monotone Boolean circuits in the initial conditions of the system, with the outcome at a particular site representing the circuit's output. Thus these systems are **P**-complete in $d \geq 3$. Below, we will give the same kind of proof for Sandpile Prediction.

In two dimensions, this proof fails for the following reason: planar circuits are **P**-complete, and monotone circuits are **P**-complete, but circuits that are both planar and monotone can be solved with an $\mathbf{NC}^3$ algorithm [28, 6, 9]. While we have not been able to adapt this algorithm to these systems except in certain special cases, the exciting possibility is open that there is a fast algorithm for these systems, including sandpiles, when $d = 2$.

[3] The reduction itself has to be computationally easy, i.e. in $\mathbf{NC}^1$ or even **DLOGTIME**. This is true for our reductions, but we will neglect to prove it.

## 3 The Abelian Sandpile

The Bak-Tang-Wiesenfeld sandpile model [1, 2] consists of a lattice where each site $i$ has a non-negative number of sand grains $s_i$. If $s_i$ is greater than or equal to a threshold equal to $s$'s coordination number $k$, then we reduce $s_i$ by $k$ and add one grain to $s_j$ for each of $i$'s neighbors $j$, so that sand is locally conserved. This is called *toppling*.[4] Toppling $i$ may raise $s_j$ above the threshold for some $j$, which may in turn set off one of $j$'s neighbors, and so on. Sand falls off the lattice at the boundary; we can think of this as having additional sites outside the lattice that act as sinks. This *avalanche* continues until we reach a stable state in which all sites are below the threshold.

In this paper, we will restrict ourselves to $d$-dimensional cubic lattices, in which $k = 2d$. However, sandpiles can be considered on any directed graph.

This model is *Abelian* in the sense that if we define an operator $\alpha_i$ that adds a grain at a site $i$ and completes the ensuing avalanche, if any, then $\alpha_i$ and $\alpha_j$ commute for all $i$ and $j$. Thus it does not matter in what order we topple the sites. In fact, if we define the sum of two stable states as the final state reached from their sum, then the set of stable states forms a commutative semigroup [7].

Sandpiles are motivated partly by a desire to find simple systems that exhibit $1/f$ noise similar to that seen in earthquakes, extinction events, the stock market, and so on. Such power-law fluctuations often occur in statistical systems, but only when the system's parameters are tuned exactly to a critical state or phase transition. Sandpiles, on the other hand, exhibit *self-organized criticality*, meaning that the dynamics tends toward the critical state from a wide variety of initial conditions and parameter settings.

A typical numerical experiment on sandpiles might consist of taking an initial state, adding a grain of sand to it at a random site, and asking what final state it reaches and what avalanches it takes to get there. By repeating this process many times, we put the system into a self-organized critical state, and can measure power laws in quantities such as the mass of the avalanche (the total number of topplings), its area (the number of sites that topple at least once) and so on.

Since the difference between two states can be written in at most one way as a sum of topplings, the set of topplings is uniquely determined by the initial and final states. Thus we can focus on the following: given the initial state of a sandpile on a $d$-dimensional lattice of finite extent, with some sites above the threshold, what final state does it reach? We call this problem SANDPILE PREDICTION.[5]

---

[4] Note that $s_i$ is actually more like the gradient in a sandpile than the number of grains of sand.

[5] We could define a more general problem which takes a description of the sandpile's lattice as input, rather than fixing it in advance. But this seems less relevant to the work people actually do.

## 4 P-completeness in $d \geq 3$

A common way to run a sandpile model is with a parallel update or cellular automaton. At each tick of a clock, we update all sites simultaneously:

$$s_i \to s_i - \begin{cases} k \text{ if } s_i \geq k \\ 0 \text{ if } s_i < k \end{cases} + \sum_{\text{neighbors } j} \begin{cases} 1 \text{ if } s_j \geq k \\ 0 \text{ if } s_j < k \end{cases}$$

It is easy to show that if the initial conditions have $s_i < 2k$ for all $i$, then this will continue to be true, so we can think of this as a cellular automaton with $2k$ states.

To show that Sandpile Prediction is **P**-complete in $d \geq 3$, we first need to show that it is in **P**. We will do this by proving that the total number of avalanches needed to relax a sandpile is polynomial in the system size.

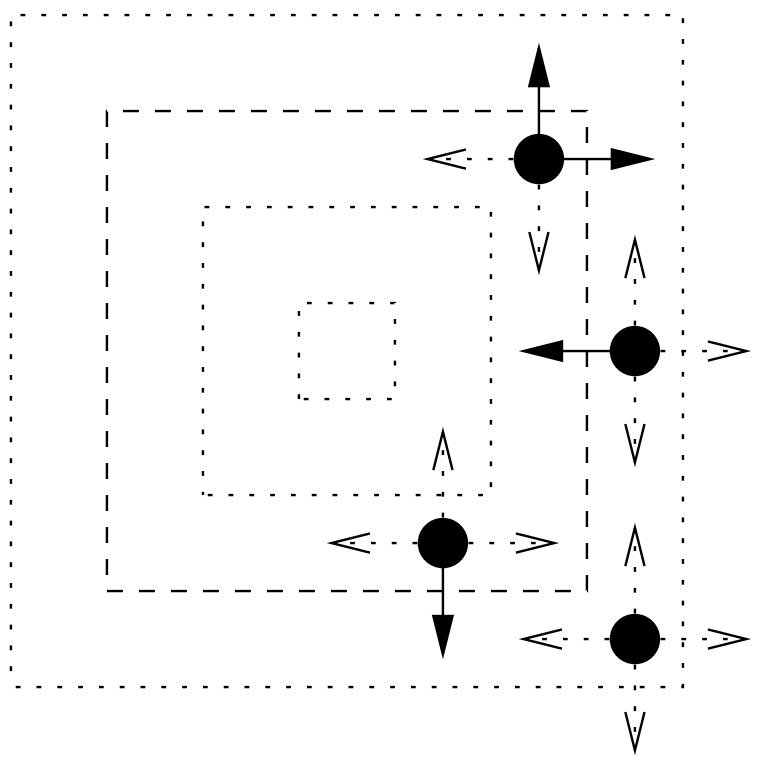

**Fig. 1.** The flux outward from one shell to the next. Topplings send at least one grain out and at most one grain in, so the flux is at least $\tau_i - \tau_{i+1}$.

Consider a $d$-dimensional cubic lattice with $l$ sites on a side where $l$ is odd, and separate it into a series of $j$ cubic shells of size $1, 3, 5, \ldots, 2j-1 = l$. A toppling in a given shell sends at least one grain into the next shell out and at most one into the next shell in, as shown in figure 1. Therefore, the flux of sand outward from the $i$th to the $(i+1)$st shell is at least $\tau_i - \tau_{i+1}$ where $\tau_i$ is the number of topplings in the $i$th shell.

Moreover, the $i$th shell has a volume of $(2i-1)^d$ and the initial state has at most $2k - 1 = 4d - 1$ grains at each site, so the flux out of the $i$th shell can be at most $(4d-1)(2i-1)^d$ before all the sand has flowed out of it (which in fact never happens). Therefore, we have

$$\tau_i - \tau_{i+1} \leq (4d-1)(2i-1)^d \text{ for all } i \geq j$$

where by convention $\tau_{j+1} = 0$. Then the total number of topplings is

$$\begin{aligned}\sum_{i=1}^{j} \tau_i &= \sum_{i=1}^{j} \sum_{i'=i}^{j} \tau_{i'} - \tau_{i'+1} \\ &\leq (4d-1) \sum_{i=1}^{j} \sum_{i'=i}^{j} (2i'-1)^d \\ &< (4d-1)\, \frac{j(j+1)}{2}\, (2j-1)^d \\ &= (4d-1)\, \frac{(l+1)(l+3)}{8}\, l^d\end{aligned}$$

This is $\mathcal{O}(l^{d+2})$, so we have

**Lemma 1.** *The total number of topplings it takes a $d$-dimensional cubic sandpile of size $l$ to relax is proportional to at most $l^{d+2}$. Therefore,* Sandpile Prediction *is in* **P**.

This bound is tight: the number of avalanches it takes a one-dimensional lattice of $l$ twos to relax is $l/6 + l^2/4 + l^3/12$ for even $l$, and in two dimensions an $l \times l$ square of fours takes about $(l/2)^4$ avalanches. Clearly, a similar bound will apply to any finite-dimensional lattice.

We will now give our **P**-completeness proof, which is shown graphically in figure 2. A wire (or rather a fuse, since it can only be used once) consists of a path of sites one grain below the threshold. If any site along the path reaches the threshold and topples, it sets off the site adjacent to it, and so on. Truth and falsehood correspond to toppling or not toppling the wire.

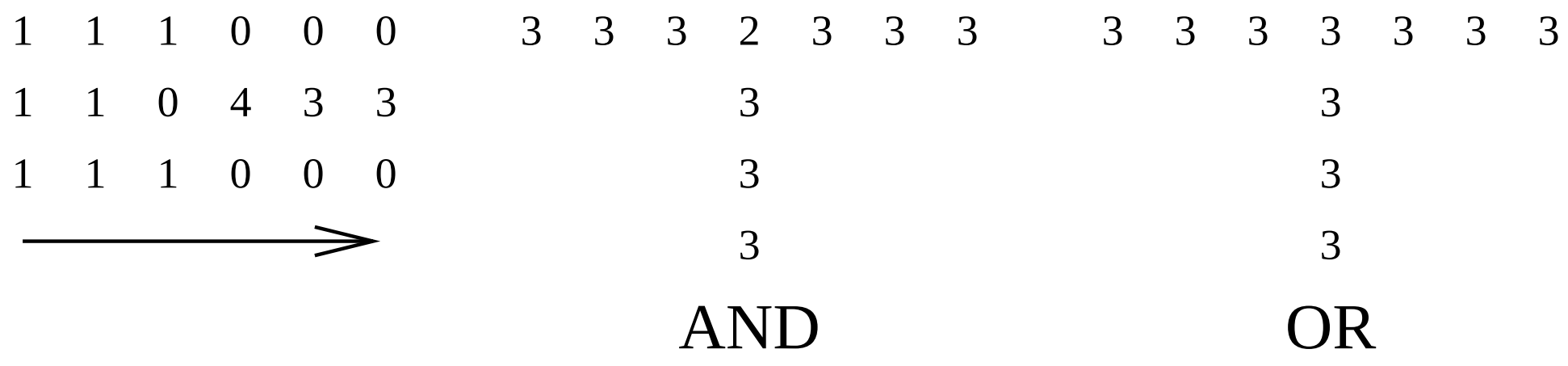

**Fig. 2.** A wire with a signal moving to the right, an AND gate, and an OR gate. In $d$ dimensions, 2, 3 and 4 should be replaced by $2d-2$, $2d-1$, and $2d$ respectively. All sites not shown are zeroes.

AND and OR gates then consist of places where three such paths meet, where the central site is two grains or one grain below the threshold respectively. If any two paths coming into an AND gate topple, the center site topples and sets off the third path; if any one path coming into the OR gate topples, then the center site topples and sets off the other two. OR gates can also be used to branch a wire into multiple copies.

These gates are essentially embeddings of the gates constructed by Bitar, Goles and Margentsten in [3, 11], who show that sandpiles on trees can implement Boolean circuits. Using a double-wire logic where each variable corresponds to a pair of wires carrying $x$ and $\overline{x}$, they implement negation as well by crossing the two wires. Since this violates planarity, negation does not appear to be possible in two dimensions.

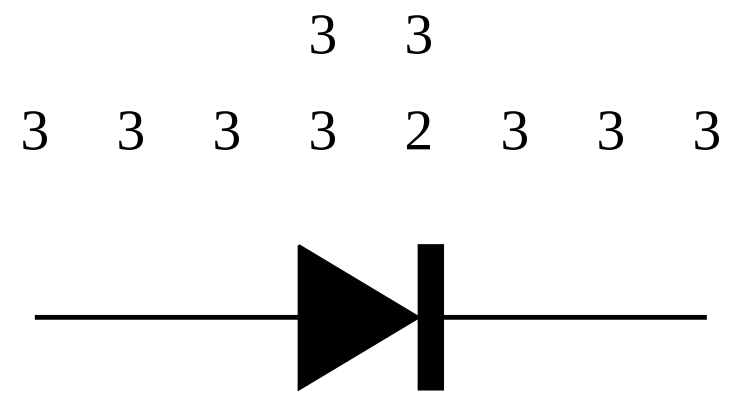

**Fig. 3.** A diode. Truth can flow from left to right, but not vice versa.

As pointed out in [23], we also need diodes to prevent signals from propagating back up our circuit the wrong way and producing spurious truths. We can build a diode by linking an OR gate to an AND gate as shown in figure 3.

In three or more dimensions, wires can cross over each other and connect gates in arbitrary ways. Moreover, the size of the lattice needed to carry this construction out is polynomial in the size of the circuit. Therefore, we have reduced MONOTONE CIRCUIT VALUE to SANDPILE PREDICTION for any $d \geq 3$, and we can state

**Theorem 2.** SANDPILE PREDICTION *is* **P**-*complete in* $d \geq 3$.

## 5 Fast algorithms for $d = 1$

As one might imagine, the one-dimensional sandpile is considerably easier to predict than its higher-dimensional cousins. However, its dynamics are non-trivial. In this section, we give serial and parallel algorithms that run in $\mathcal{O}(n \log n)$ and $\mathcal{O}(\log^3 n)$ time respectively, thus showing that SANDPILE PREDICTION is in $\mathbf{NC}^3$ for $d = 1$.

We note that lower bounds for the size of avalanches, and a closed-form solution for a special case where $n$ grains of sand are piled at a single site, are given in [10].

Our initial states will consist of zeroes, ones and twos; we can add 3's and higher values if we like. Since the time evolution of the sandpile is Abelian, we can separate the relaxation process into one avalanche for each 2, and handle them in whichever order we please. For instance, the state 00121200 is 00121100 + 00000100, so we can first relax from 00121100 to 01110110, and then add the

second 2, giving 01110210 which avalanches to the final state 01111101. We are simply taking the incrementing operators $\alpha_i$ in series.

The outcome of the avalanche from a single two in a block of ones is easy to determine. It travels out in both directions, with left- and right-moving boundaries $\cdots 11120111 \cdots$ and $\cdots 11102111 \cdots$, until it hits the closest zeroes, covers them with a grain each of sand, and travels back inwards. Finally, it leaves a single zero somewhere inside the original block. The position of the final zero can be determined by noting that the center of mass $\sum_x x\, s_x$ is conserved, since each toppling sends one grain each to the left and right.

We can subtract a mass of 1 from each site, so that twos and zeroes have a mass of $+1$ and $-1$ respectively. Then if we have a two at $t$ and zeroes at $z_1$ and $z_2$, the center of mass is $t - z_1 - z_2$ so the final zero will be at $z' = z_1 + z_2 - t$.

Here is an example:

0111121110
0111202110
0112020210
0120202020
0202020201
1020202011
1102020111
1110201111
1111011111

The initial two is at 6, the zeroes are at 1 and 10, and the final zero is at $1 + 10 - 6 = 5$.

This suggests the following serial algorithm. Make a list $Z = \{z_1, z_2, \ldots\}$ of the positions of zeroes in the initial state from left to right, and a similar list $T = \{t_1, t_2, \ldots\}$ of the positions of the twos. Take the twos in turn. If $t_i$ is also the position of a zero, then incrementing that site simply cancels the zero there, so remove it from both $T$ and $Z$. Otherwise, do a binary search in $Z$ to find the closest zeroes such that $z_j < t_i < z_{j+1}$, replace $z_j$ and $z_{j+1}$ with $z_j + z_{j+1} - t_i$, and remove $t_i$ from $T$. Continue through $T$ until there are no twos left. The final list $Z$ will be the positions of zeroes in the final state, in a background of ones.

For the initial state 00202120110, for instance, we have $Z = \{1, 2, 4, 8, 11\}$ and $T = \{3, 5, 7\}$ and the algorithm transforms $Z$ as follows:

$$\{1,2,4,8,11\} \stackrel{\text{two at } 3}{\longrightarrow} \{1,3,8,11\} \stackrel{\text{two at } 5}{\longrightarrow} \{1,6,11\} \stackrel{\text{two at } 7}{\longrightarrow} \{1,10\}$$

so that the final state is 01111111101, which the reader can easily verify.

To handle the standard boundary conditions in which the sites outside the lattice at 0 and $n+1$ act as sinks, we add a sufficient quantity of zeroes there, one for each two, so that they can never be filled up. For instance, the state 112120 becomes $Z = \{0, 0, 6, 7, 7\}$ and $T = \{3, 5\}$, and our algorithm yields a final state of $Z = \{0, 5, 7\}$ or 111101 with one zero left at each end.

On a lattice of size $n$, this algorithm will take at most $n$ steps since there are at most $n$ twos. Each step involves a binary search through at most $n$ zeroes, and addition and subtraction of integers with $\mathcal{O}(\log n)$ digits. Since both of these things can be done in $\mathcal{O}(\log n)$ time, we have

**Theorem 3.** Sandpile Prediction *can be solved in* $\mathcal{O}(n \log n)$ *serial time in* $d = 1$.

It is not obvious that this algorithm can be efficiently parallelized. If two twos create non-overlapping avalanches, certainly we can calculate their outcomes in parallel; but how do we know in advance how various avalanches will overlap? To construct our parallel algorithm, we discuss two other problems first, Reachability and Generability.

The input to Reachability consists of a directed graph with $n$ nodes, a start vertex $a$ and an end vertex $b$. Is there a path from $a$ to $b$? At first glance, this requires us to search throughout the graph, which might take $n^2$ steps. However, by squaring the graph's adjacency matrix $\mathcal{O}(\log n)$ times we can obtain its transitive closure, thus telling us whether $b$ is reachable from $a$ for all $a$ and $b$. Since squaring an $n \times n$ Boolean matrix can be done in $\mathcal{O}(\log n)$ parallel steps, Reachability can be solved in $\mathcal{O}(\log^2 n)$ parallel time, in $\mathbf{NC}^2$ [26].

Generability[6] consists of a set $W$ with $n$ elements, a binary operation $\bullet$ from $W \times W$ to $W$ which may be partial or multivalued, an initial subset $V \subset W$, and a target element $w$. Is $w$ generable from $V$? That is, by starting with elements in $V$ and repeatedly combining them with $\bullet$, can we obtain $w$? This is different from Reachability in a simple but important way: two elements combine to make a third, rather than single elements generating each other through arrows. This is enough to make Generability **P**-complete [12].

However, consider the following special case. Suppose that every element $w \in W$ has a positive integer *rank* $r(w)$ which is additive or superadditive, $r(u \bullet v) \geq r(u) + r(v)$. Suppose also that the largest rank, $\max_{w \in W} r(w)$, is polynomial in $n$. We call this Additive Ranked Generability.

**Lemma 4.** Additive Ranked Generability *is in* $\mathbf{NC}^3$.

*Proof.* We will reduce Additive Ranked Generability to a series of $\mathcal{O}(\log n)$ Reachability problems. Assume without loss of generality that $r(v) = 1$ for all $v \in V$.

Define $V_0 = V$, and generate a series of sets $V_i$ inductively as follows. Given $V_{i-1}$, define a directed graph with arrows $w_1 \to w_2$ if $w_2 = w_1 \bullet v$ or $w_2 = v \bullet w_1$ for some $v \in V_{i-1}$. Then define $V_i$ as all the elements reachable from $V_{i-1}$, including those already in $V_{i-1}$.

We will now show that $V_i$ is exactly the set of elements generable from $V$ of rank less than or equal to $2^i$, or equivalently, that if $w$ is generable from $V$, that it appears in $V_i$ for $i = \lceil \log_2 r(w) \rceil$. This is true for $i = 0$ by definition, since the only generable elements of rank 1 are those in $V_0 = V$ itself. Assume it is true for all elements with rank less than $r(w)$.

[6] Also called Path System [12].

If $w$ is generable, $w = u \bullet v$ where $u$ and $v$ have smaller rank. Assume that $r(u) \geq r(v)$. Then $r(v) \leq r(w)/2$, and by the induction hypothesis, $v \in V_{i-1}$ since $\lceil \log_2(r(w)/2) \rceil = \lceil \log_2 r(w) \rceil - 1$. Also, $u \in V_i$ by induction since $u$ has smaller rank than $w$. But by definition, $V_i$ contains all elements that can be written $u \bullet v$ where $u \in V_i$ and $v \in V_{i-1}$, so $w \in V_i$. This proves the induction.

So $V_i$ contains all generable elements of rank up to $2^i$. Since the maximum rank of any element is polynomial in $n$, all generable elements are in $V_{\mathcal{O}(\log n)}$, and we only need to take $\mathcal{O}(\log n)$ steps calculating $V_i$ from $V_{i-1}$. Each one of these is a REACHABILITY problem, and can be done in $\mathcal{O}(\log^2 n)$ parallel time.[7] Thus the total time is $\mathcal{O}(\log^3 n)$, or in $\mathbf{NC}^3$. □

To predict the one-dimensional sandpile, we wish to identify the smallest regions whose avalanches do not overlap, and so which do not interact with each other. We call these regions *chunks*, and associate them with the set of twos they contain. Their final state has a simple form:

**Lemma 5.** *The final state of any chunk consists of either a strip of ones, or a strip of ones with a single zero.*

*Proof.* Since the system is Abelian, we can add the twos one at a time. The simplest chunk consists of two zeroes with a single two at $t_1$ between them, and we showed above that its final state has a single zero at $z' = z_1 + z_2 - t_1$. If we then add another two at $t_2$ on either side of $z'$, the avalanche covers both the zero at $z'$ and another one farther away at $z_3$, leaving a larger block of ones with a new zero at $z'' = z' + z_3 - t_2$. If a two falls directly on the zero, it cancels it and leaves us with a block of ones. Either way, we are left with one zero, or none, in a block of ones. □

All we have to do now is find the boundaries of these chunks, since the final state of each one is easily determined by conserving the center of mass. Define a *possible chunk* as a set of twos which, if no other twos were present, could evolve to a strip of ones with at most one zero, without violating conservation of mass or center of mass. (In other words, we imagine initial states with all of the zeroes present, but only some of the twos.) Define the *extent* of a possible chunk as $[z_1, z_2]$ where $z_1$ and $z_2$ are the leftmost and rightmost zeroes that the avalanches of those twos would cover.[8] Possible chunks are easy to find:

**Lemma 6.** *Given a set of twos, we can determine if they form a possible chunk, and if so, its extent and final state, in $\mathcal{O}(\log n)$ parallel time.*

---

[7] In fact, since REACHABILITY can be solved in $\mathbf{NL}$, nondeterministic logarithmic space, we can say that ADDITIVE RANKED GENERABILITY is in $(\mathbf{NC}^1)^{\mathbf{NL}}$, i.e. it can be solved in $\mathcal{O}(\log n)$ parallel time with an $\mathbf{NL}$ oracle.

[8] Not all possible chunks are actual, since conservation of mass and center of mass are not enough in themselves to determine the sandpile's dynamics. For instance, 022200002220 could evolve to 12 ones without violating the conservation laws, but it actually consists of two chunks and evolves to 110111111011 after spilling one grain off each end of the lattice.

*Proof.* Suppose we have initial conditions with $n$ zeroes at $z_1, z_2, \ldots, z_n$ (increasing from left to right) and $m$ twos at $t_1, t_2, \ldots, t_m$ with $z_1 < t_1$ and $z_n > t_m$, and ones at all other sites between $z_1$ and $z_m$. If this is a chunk, then either the final state is all ones, in which case $n = m$ and both the mass and the center of mass (minus one for each site) is zero, and

$$\sum_i z_i - \sum_i t_i = 0 \tag{1}$$

or the final state has a single zero, $n = m + 1$ so the mass is $-1$, and

$$z_1 < z' < z_n \text{ where } z' = \sum_i z_i - \sum_i t_i \tag{2}$$

Then this set of twos and zeroes forms a possible chunk with extent $[z_1, z_n]$ if and only if (1) or (2) is true, and the final state in the latter case is given by $z'$.

To complete the proof, we need to show that the extent is well-defined; that is, that for any given set of twos, that there is only one pair $[z_1, z_n]$ such that they and the zeroes between them form a chunk with these twos. If not, we have three cases. If equation (1) holds with two different pairs $[z_1, z_n]$ and $[z_{k+1}, z_{n+k}]$ for some $0 < k < n - 1$, subtracting them yields

$$\sum_{i=1}^{k} (z_{n+i} - z_i) = 0$$

which is a contradiction since every term in the sum is positive. If (2) holds with two different pairs $[z_1, z_n]$ and $[z_{k+1}, z_{n+k}]$ with final zeroes at $z_1 < z' < z_n$ and $z_{k+1} < z'' < z_{n+k}$ respectively, subtracting yields

$$\sum_{i=1}^{k} (z_{n+i} - z_i) = z'' - z'$$

so

$$z'' \geq z_{n+k} + (z_{n+k-1} - z_k) + (z_{n+k-2} - z_{k-1}) + \cdots + (z_n - z_2) + (z' - z_1)$$

Since $n > 1$ and $z' > z_1$, every term in the sum is positive, so $z'' > z_{n+k}$, another contradiction. Finally, if (1) and (2) hold for pairs $[z_1, z_n]$ and $[z_k, z_{n+k}]$ respectively with a zero at $z'$ in the latter case, subtracting yields

$$\sum_{i=1}^{k} (z_{n+i} - z_i) = z' - z_k$$

If $k \geq 1$ this leads to the same kind of contradiction as before, namely that $z' \geq z_{n+k}$. If $k < 1$ we switch left and right and a symmetrical argument shows that $z' \leq z_k$. Thus at most one equation of the form (1) or (2) can hold.

These sums involve adding at most $n$ numbers with at most $\log n$ digits, which can be carried out in $\mathcal{O}(\log n)$ parallel time [26]. $\square$

Finally, we can explain our prediction algorithm. First, consider all contiguous subsets of the set of twos. Apply lemma 6 to all of them in parallel, to determine which ones are possible chunks and what their extent is. Finding which possible chunks are actual then becomes a GENERABILITY problem: if two neighboring chunks have overlapping extent, they will combine into a single larger chunk.

Formally, let $W$ be the set of $\binom{n+1}{2}$ contiguous sets of twos $\{t_i \cdots t_j\}$ for $j \geq i$. Define $\{t_i \cdots t_j\} \bullet \{t_k \cdots t_l\} = \{t_i \cdots t_l\}$ if both are possible chunks with extents $[z_p, z_q]$ and $[z_r, z_s]$ respectively, if $z_r \leq z_q$, and if $j = k-1$. Let $\bullet$ be undefined in all other cases. Define the rank as the number of twos, $r(\{t_i \cdots t_j\}) = j - i + 1$. The rank is then additive, and its maximum is linear in the system size.

Then the actual chunks are the generable ones, and the maximal generable chunks are the maximal parts of the system that are non-interacting. We can identify all generable chunks in $\mathbf{NC}^3$ with lemma 4, and pick out the maximal ones in $\mathcal{O}(\log n)$ time. Finally, we use lemma 5 to compute the final state within each chunk, and we're done. Thus we have proved

**Theorem 7.** SANDPILE PREDICTION *in* $d = 1$ *can be reduced to* ADDITIVE RANKED GENERABILITY. *Thus it can be solved in* $\mathcal{O}(\log^3 n)$ *time and is in* $\mathbf{NC}^3$.

We can derive the exact set of topplings that take place from the initial and final sandpile states $s_{\text{init}}$ and $s_{\text{fin}}$ as follows. Consider the *toppling matrix*

$$T = \begin{pmatrix} -2 & 1 & & & \\ 1 & -2 & 1 & & \\ & 1 & -2 & & \\ & & & \ddots & \\ & & & & -2 \end{pmatrix}$$

where $T_{ij}$ is the number of grains added to site $j$ when site $i$ topples. Its inverse has a simple form:

$$(T^{-1})_{ij} = -\frac{(n+1-i)\,j}{n+1} \text{ for } i \geq j$$

and symmetrically when $j \geq i$. Then $T^{-1}(s_{\text{fin}} - s_{\text{init}})$ gives a vector whose $i$th component is the number of times the $i$th site toppled.

If we are interested just in the mass of an avalanche, i.e. its total number of topplings, we can define a quantity

$$E(s) = (1/2) \sum_x x^2 \, s_x$$

An equivalent quantity is called the *energy* in [10]. It is easy to see that each toppling reduces $E$ by 1 (if we keep track of sand that falls off the boundaries), so the total number of topplings is simply $E(s_{\text{init}}) - E(s_{\text{fin}})$.

## 6 The two-dimensional case

There is a great deal of theoretical and numerical interest in the two-dimensional sandpile, so a fast parallel algorithm for it would be both exciting and useful. What are the prospects for such an algorithm, or for a proof of **P**-completeness?

In two dimensions, our fast algorithm for $d = 1$ fails because there are an exponential, rather than polynomial, number of possible chunks, since there are an exponential number of two-dimensional blobs of a given area. Therefore, this algorithm shows little hope of generalizing.

On the other hand, our **P**-completeness proof for $d = 3$ fails because Boolean circuits that are both monotone and planar can be evaluated with an algorithm in $\mathbf{NC}^3$ [28, 6, 9]. This looks hopeful, but two-dimensional sandpiles differ from planar Boolean circuits in two ways. First, connections between sites are bidirectional. Secondly, a given site can change a polynomial number of times in the course of the sandpile's evolution. Thus we really have a three-dimensional Boolean circuit of polynomial depth, with layers corresponding to successive steps in the sandpile's space-time.

The reader may be able to use this space-time picture to find a clever embedding of non-planar Boolean circuits, thus rendering the $d = 2$ case **P**-complete. However, the Abelianness of sandpile dynamics seems to prevent wires from crossing. Alternately, she may be able to show that the simple periodic structure of this three-dimensional circuit allows it to be evaluated quickly.

A third possibility is that two-dimensional sandpiles are not in **NC**, but are not **P**-complete either. Such problems are known to exist if $\mathbf{NC} < \mathbf{P}$, but only artificial ones constructed for this purpose [27]. This would be a frustrating possibility, since there are currently no proof techniques to place a natural problem in this gap.

## 7 Recurrence and the identity

A sandpile state is *recurrent* if any state can be transformed to it by adding sand in the right places and letting the system relax. Let us call Sandpile Recurrence the problem of determining whether a given state is recurrent. In $d = 1$ this problem is trivial, since recurrent states are those with one or no zeroes. How hard is it in higher dimensions?

If we think of all boundary sites as connected to a single sink by the number of edges leaving the lattice from each one, then toppling the sink adds the corresponding number of grains to each boundary site. On a square lattice, for instance, we would add one grain to each edge site and two to each corner. Majumdar and Dhar [19] showed that a state is recurrent if and only if every site topples in the avalanche that follows from this; this is often called the "burning algorithm." They went on to show that there is a one-to-one correspondence between recurrent states and spanning forests rooted at the boundary, where we draw an edge between each site and the neighbor responsible (according to an arbitrary ordering) for its toppling.

Using the burning algorithm, we can show that Sandpile Recurrence is **P**-complete in $d \geq 3$. We start by filling the entire lattice with $k - 1 = 2d - 1$ grains of sand per site. This is clearly recurrent since any state can be filled up to this level by adding more sand. We then burrow through the lattice, building a circuit with reduced amounts of sand, so that the sites in the circuit will be left with the right amount after every site outside the circuit topples. Wires should be built with 1 grain per site, so that they have $k - 1$ grains after the $k - 2$ sites around them topple; AND and OR gates have $k - 3$ neighbors outside the circuit, so they should start with 1 and 2 grains each in order to end up with $k - 2$ and $k - 1$ respectively. Finally, sites corresponding to the circuit's inputs are given one grain if true and no grains if false.

We wish to have every site in the circuit topple if the circuit's output is true. We can do this by branching the output with OR gates and feeding a copy of it back onto every wire in the circuit, with diodes to prevent truth from flowing the wrong way. Then every wire becomes true if and only if the output does, and we have proved

**Theorem 8.** Sandpile Recurrence *is* **P**-*complete in* $d \geq 3$.

If we denote by $a + b$ the result, after avalanches, of adding two states $a$ and $b$ site-by-site, then by definition $c$ is recurrent if for all $a$, there exists a $b$ such that $c = a + b$. Since $+$ is both commutative and associative, it follows that the set of recurrent states forms an Abelian group, and that it must have an identity $e$ such that $a + e = a$ for all recurrent states $a$.

In $d = 1$, the identity on a lattice of size $n$ consists of ones, with a zero in the middle if $n$ is odd. In higher dimensions the identity is a complicated and beautiful pattern; the identity for a $511 \times 511$ lattice is shown in figure 4. The only regularities known about the $n \times n$ identity seem to be that if $n$ is even it has a large block of twos in the center, and if $n$ is odd it consists of the $(n-1) \times (n-1)$ identity plus a central cross [8]. Therefore, we propose Sandpile Identity as another problem worthy of study.

The fastest algorithm we have been able to find in the literature [5] consists of taking the non-recurrent state that results from toppling the sink onto the zero state (for a rectangular lattice, one grain on each edge site, two on the corners, and zero in the interior) and repeatedly doubling it until we reach a recurrent fixed point. Since the number of doublings required is logarithmic in the volume of the lattice, this can be done in **NC** if Sandpile Prediction can; but since this a single special case, there may be a fast algorithm even when general prediction is **P**-complete. Perhaps a clever reader can find a faster or more direct construction.

## 8 Conclusion

As in other discussions of **P**-completeness in statistical physics, this paper begs the following question: How relevant is it that the microstate of a sandpile, in the worst case, is difficult to predict? Aren't physicists interested more in bulk properties than in microstates, and in average cases more than in worst ones? This is analogous to the question of whether the **P**-completeness of lattice gases really says anything about the complexity of the Navier-Stokes equations [22].

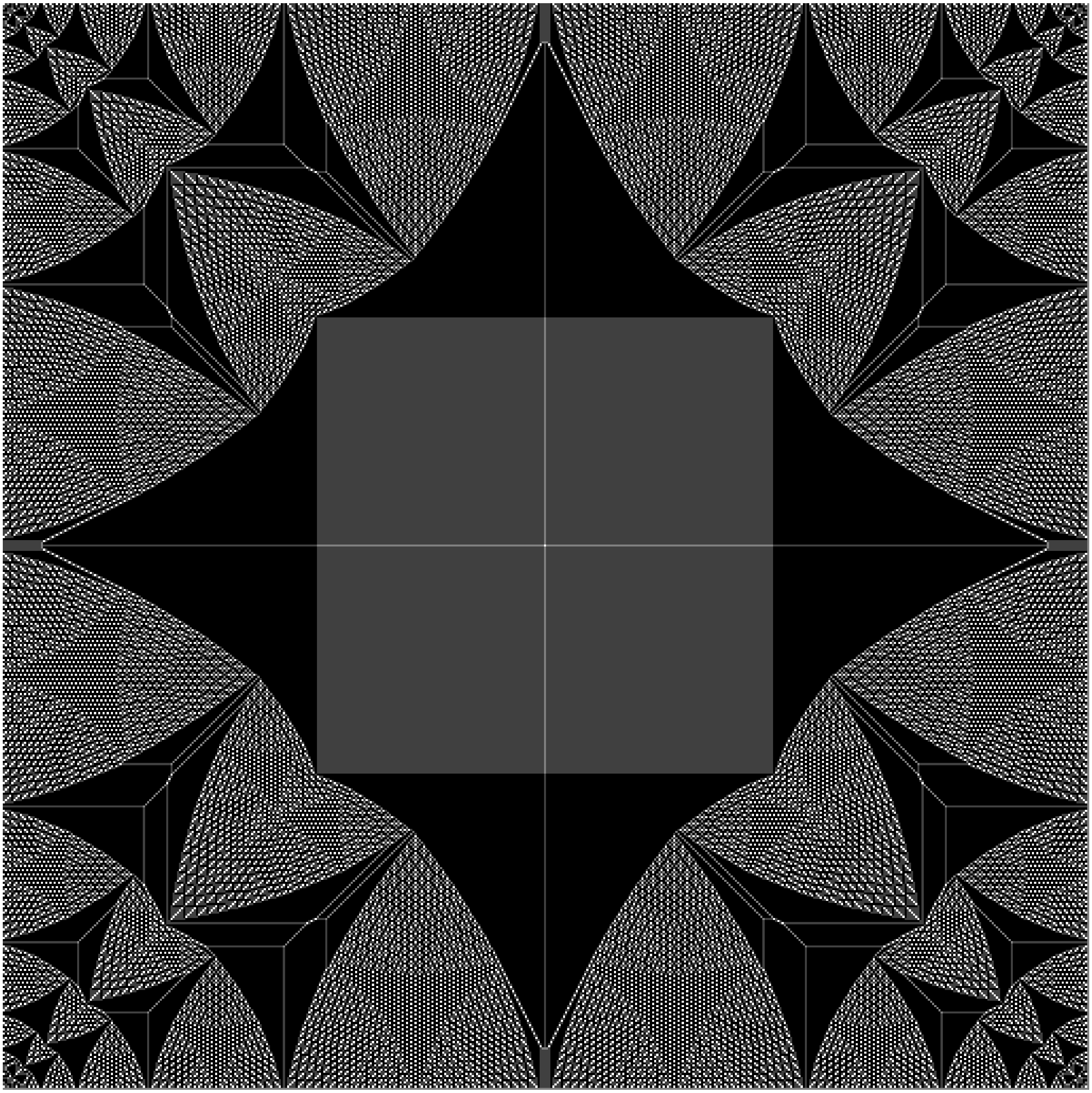

**Fig. 4.** The identity configuration for a $511 \times 511$ lattice. Black is three grains per site, white is zero.

This seems to us to be an important area for further research. Computer scientists are working to define average-case complexity, and many **NP**-complete problems are actually easy most of the time. However, we still feel that **P**-completeness shows that a system is highly contingent on its own history, and that this is a fundamental qualitative feature of its dynamics.

Even if $\mathbf{NC} < \mathbf{P}$, there might still be a parallelization of the problem that achieves a polynomial speedup, predicting $t$ time-steps in $\mathcal{O}(t^\alpha)$ parallel steps for some $\alpha < 1$. Moriarty, Machta and Greenlaw give such an algorithm for diffusion-limited aggregation [25]. The theory of *strict* **P**-*completeness* can be used to put lower bounds on $\alpha$, assuming that there are **P**-complete problems for which no such speedup is possible [4, 22].

Finally, we note that the one-dimensional sandpile is a special case of particle ballistics in cellular automata, where solitons such as $\cdots 11102111 \cdots$ travel, collide, annihilate and produce new particles according to a certain 'chemistry' (e.g. [14]). Here, conservation of mass and center of mass allows us to greatly simplify the problem, but there may be other interesting cases which are also efficiently predictable.

**Acknowledgements.** We wish to thank Cosma Shalizi for preparing figure 4 on the CAM-8. C.M. would like to thank Molly Rose and Spootie the Cat for enlightenment and support.